\documentclass[]{article}
\usepackage{emulateapj}
\usepackage{psfig}
\usepackage{graphicx}

\advance \voffset by 0.00cm\relax
\def\msun{{\rm ~M}_{\odot}}
\def\rsun{{\rm ~R}_{\odot}}

\def\mpy{{\rm ~M}_{\odot} {\rm ~yr}^{-1}}
\begin{document}

\title{GALACTIC POPULATIONS OF ULTRACOMPACT BINARIES}

\author{Krzysztof Belczynski\altaffilmark{1,2,3}, Ronald E. Taam\altaffilmark{1}}

\affil{
     $^{1}$ Northwestern University, Dept. of Physics \& Astronomy,
       2145 Sheridan Rd., Evanston, IL 60208\\
     $^{2}$ Nicolaus Copernicus Astronomical Center,
       Bartycka 18, 00-716 Warszawa, Poland;\\   
     $^{3}$ Lindheimer Postdoctoral Fellow\\  
     belczynski, r-taam@northwestern.edu}

\begin{abstract} 
Recent {\em RXTE} and {\em Chandra} discoveries of low mass X-ray 
binaries with ultra-short orbital periods have initiated theoretical 
work on the origins of these peculiar systems.  
Using the {\em StarTrack} population synthesis code the formation and 
evolution of X-ray ultracompact binaries (UCBs) in the Galactic field are 
analyzed.
The relative number of UCBs with a neutron star or a 
black hole accretor populating our Galaxy is predicted.  Our results 
demonstrate that standard evolutionary scenarios involving primordial 
binaries can be sufficient to produce the UCBs in the Galactic field
without requiring additional processes associated with the dense 
stellar environments in the cores of globular clusters.   
In contrast to previous studies we find that the 
majority of the immediate progenitors of these systems consist of a 
hydrogen exhausted donor with an ONeMg white dwarf. The evolution of these  
systems leads to the accretion induced collapse of the white dwarf to 
a neutron star, which can  play an important role in the formation of a 
majority of Galactic UCBs. We predict that with an increase in the number of 
X-ray active UCBs hosting neutron stars by an order of magnitude, a system 
with a black hole accretor may be found. 
\end{abstract}


\section{INTRODUCTION}

Recent discoveries have revealed that the X-ray binary population 
of the Galaxy extends to orbital periods as short as 20 
minutes. Their existence in both the Galactic field and in the 
dense stellar environments in globular cluster systems has motivated 
theoretical work on their formation and evolution. The class of 
previously known ultra-short period systems known as AM CVn systems 
are generally believed to consist of a white dwarf (WD) accreting 
material from another degenerate white dwarf or from a low mass 
semi-degenerate companion (for a review, see Warner 1995). In contrast,  
the X-ray emitting systems, some of which are millisecond pulsars, contain a 
neutron star (NS) accretor and, most probably, a degenerate low mass donor.
To date, only 4 systems are confirmed while another 4 are strong candidates 
of UCBs with a NS accretor in the Galactic field (Chakrabarty 2003). So far, 
no system with a black hole (BH) accretor has been discovered.

There are 
three proposed scenarios of UCB formation, 
involving the evolution of a NS-WD or NS-He star 
binary (descendant of a common envelope episode), or the mass
transfer (MT) from an evolved main sequence (MS) donor to a NS. 
Although the ultracompact phase has been studied in detail (Tutukov et al. 1987; 
Rasio, Pfahl \& 
Rappaport 2000; Yungelson, Nelemans, \& van den Heuvel 2002; Nelson \&
Rappaport 2003; Deloye \& Bildsten 2003), the prior evolution of progenitor 
systems was usually assumed or only briefly discussed in different contexts 
(e.g. Nelemans, Yungelson \& Portegies Zwart 2001). 
Our analysis of possible evolutionary channels leading to the formation 
of the UCBs with NSs and possibly BHs shows the limitations of such an approach.
We point out in the following sections that a large fraction of NS-WD systems 
evolve first through NS-He star MT phase, with a NS forming predominantly 
via an accretion induced collapse (AIC) of a white dwarf.  In addition, we show 
that the NS-MS formation channel 
is less important for the overall production of UCBs.

\section{MODEL DESCRIPTION}

The calculations are based on a binary population synthesis method. 
The {\em StarTrack} code (Belczynski, Kalogera, \& Bulik 2002)
has recently undergone major revisions and updates (Belczynski,
Kalogera, Taam \& Rasio 2003, in preparation), among which include:
detailed treatment of tidal synchronization and circularization,
individual treatment of various MT phases, full 
numerical orbit evolution with angular momentum losses due to magnetic 
braking, gravitational radiation (GR), mass transfer/loss and tides,
and incorporate the stabilizing influence of optically thick winds 
on accreting white dwarfs (Hachisu, Kato, \& Nomoto 1996) allowing 
for growth of different types of WDs to either the disruption of the 
white dwarf in a Type Ia supernova explosion or to the formation of 
a NS via AIC (see Li \& van den Heuvel  
1997; Ivanova \& Taam 2003).  For the accumulation 
of helium or carbon/oxygen matter on CO or ONeMg WDs, we make 
use of the helium accumulation efficiencies determined by Kato \& Hachisu 
(1999) and the off center ignition results of Kawai, Saio, \& Nomoto (1987)
in determining the fate of the underlying WD.

A large sample of single ($10^6$) and binary ($10^6$) stars in the
Galactic field are evolved for 10 Gyrs assuming a 
constant star formation rate. The more
massive primaries of the binaries (and single stars) are drawn from 
an initial mass function with a  slope of -2.7 within the range of $4-100 \msun$,
while the less massive secondaries of the binaries are taken from a 
flat mass ratio distribution within $0.08-100 \msun$. Here, the mass 
ratio, $q$, is defined as the mass of the secondary to the primary.
The initial binary separations are taken from a distribution flat in  
the logarithm up to $10^5 \rsun$, while we assume a thermal equilibrium 
distribution for the eccentricities. All stars are evolved based on the 
results of Hurley, Pols, \& Tout (2000) for models with a solar metallicity.
We adopt parameters corresponding to the standard model of Belczynski,
Kalogera, \& Bulik  (2002), changing the maximum NS mass to $2 \msun$, 
incorporating the latest natal kick distribution of Arzoumanian, Chernoff 
\& Cordes (2002), and limiting the accretion rate onto the NS and the BH at the 
maximum Eddington limit with the rest of the transferred material lost with
specific orbital angular momentum of the accretor during the dynamically stable MT events,
but allowing for hyper-critical accretion at common envelope (CE) phases.  
Besides the dynamically unstable MT events we also allow 
for evolution into the CE phase in cases in which the trapping radius 
exceeds the Roche lobe radius of the accretor (e.g., King \& Begelman 1999; 
Ivanova, Belczynski, Kalogera, Rasio \& Taam 2003).

\section{RESULTS}

Since the entire spectrum of binaries and single stars are produced by 
our calculations we shall adopt as a working  definition for UCBs
any system in which mass is transferred to a NS or a BH at orbital periods 
less than some value.  For definiteness, 80 minutes is chosen as a customary 
value, but 5 hours is also used for a broader definition.

\subsection{Formation Channels}

The various UCB formation channels realized in our {\em StarTrack} 
evolutionary model are presented in Table 1. The majority of UCBs are 
formed with a NS accretor, in particular along the channel NS:01. A 
typical progenitor system corresponds to an intermediate mass primary and 
low mass secondary in an orbit sufficiently wide such that the primary 
evolves up to the asymptotic giant branch stage prior to filling its 
Roche lobe. The ensuing MT is dynamically unstable, leading to the formation
of a CE (see Iben \& Livio 1993; Taam \& Sandquist 2000). Systems 
emerging from the CE phase consist of an ONeMg WD  
(descendant of primary) and a MS secondary at an orbital 
separation such that the secondary can overfill its Roche lobe while
expanding in the Hertzsprung gap or the first giant branch. 
The further evolution 
leads to a second CE phase. Provided that the system avoids merger
the remnant system consists of an ONeMg WD and a low mass He star or WD 
companion. Hence, initially wide systems become very tight after the 
successful ejection of mass in the two common envelope phases.  The
secondary fills its Roche lobe for a second time due to the action of  
angular momentum losses associated with GR emission and, for sufficiently 
massive He stars, stellar expansion induced by nuclear evolution. However, 
in this Roche lobe overflow phase, the two components of the system 
are of comparable mass and the mass transfer proceeds stably.  
Provided that the mass transfer rates exceed a critical value such that the 
the accreting WD accumulates sufficient matter to exceed the Chandrasekhar 
mass, a NS may form by the AIC process as discussed by Taam \& van den
Heuvel (1986) and Webbink (1992) for the formation of low mass X-ray binary
systems.  More recently, Yungelson, Nelemans, \& van den Heuvel 
(2002) described the formation of the UCB X-ray pulsar 4U 1626-67 in terms 
of the AIC scenario. 
In general, the outcome of this formation channel is a UCB system consisting  
of a He star or WD donor transferring mass to its NS companion. 

In contrast to the NS UCBs, there is no single major formation channel for 
the BH UCB systems. Although the BH:01 dominates in the formation of BH UCBs, 
the effect is not as strong as in the case of NS UCBs with the  NS:01 channel. 
Close inspection of Table 1 reveals that the NSs in UCBs are either formed through 
AIC or Type II SN (core collapse and explosion of a massive star).
However, the BHs are not formed directly from the core of a massive star, but 
rather formed in the collapse of the accreting NS when its mass exceeds 
$\sim 2 \msun$. Therefore, the formation channels of BH UCBs without an SN 
entry, are characterized by two AIC events: a WD to a NS, and then a NS to a BH.

Most of the UCB accretors are formed through AIC of a heavy ONeMg WD to NS ($\sim$
80\%). Contrary to previous assumptions on the type of donor formed in NS-WD systems, 
the last CE episode results in the formation of not only WD ($\sim$
40\%) but also low mass He-stars ($\sim$ 40\%) secondaries. Either the WD or 
the He-star companion 
fills its Roche lobe and starts transferring material to the ONeMg WD.
The AIC interrupts the MT due to the loss of binding energy of the collapsing dwarf.
However, in the case of a He-star donor, MT may restart on a short timescale 
as nuclear expansion of the He-star is faster in bringing the system to contact 
than GR in the case of a WD donor.  The 
He-star donors eventually lose sufficient mass to become low mass ($\sim 0.35 \msun$)
hybrid WDs (with a mixture of helium
and carbon/oxygen in the core) while the systems enter 
a long-lived  ($\sim 1$ Gyr) UCB phase.
These donors, given enough time, may cool down and crystallize, forming a
Ne-enriched layer in their interiors (Yungelson, Nelemans \& van den Heuvel
2002). The subsequent 
MT eventually uncovers the deeper layers of the hybrid WD 
giving rise to a Ne-enriched accretion flow, claimed to be observed 
in several of the NS UCBs (Juett, Psaltis \& Chakrabarty 2001).
As our simulations show, a significant fraction of the NS UCBs have hybrid WD
donors.

The initial binary parameters of the systems forming UCBs with orbital 
periods less than 80 minutes are illustrated in Fig. 1.  The progenitors
of both classes of UCBs with NS and BH originate in a wide range of initial 
orbital periods and mass ratios. Upon detailed inspection, the initial 
component masses exhibit traces of correlations. For example, the 
secondary mass, for BH UCB progenitors, is confined to a narrow range
of 4 - 7 $\msun$, and although the primary mass spans a wide range, most
of the primaries have a mass close to $8 \msun$.  On the other hand, 
the progenitor systems that form NS UCBs are characterized by secondaries mostly 
within the range of 1 - 5 $\msun$, while the primaries concentrate around $7 \msun$.
It is natural that the progenitors of BH systems are more massive than the progenitors 
of NS binaries, however, contrary to expectation, the separation of the two
sub-populations is not well defined, with the two groups partially 
overlapping.  This is a direct consequence of the fact that the BHs are 
formed from the NSs via the AIC process and, thus their progenitors do not 
significantly differ from the progenitors of NS UCBs.
A few BH accretors in UCBs originate from the primaries of $\sim 30
\msun$, which is not shown in Fig. 1.

\subsection{Evolution at Ultracompact Phase}

At the short orbital periods characteristic of the UCBs, its evolution is governed by 
the mass and angular momentum loss from the system. The orbital 
evolution is dependent on the fractional amount of matter which is 
ejected from the system, angular momentum losses associated with 
this ejection of matter and GR or magnetic braking, 
as well as the possible nuclear evolution of the donor. 

An example of the evolution for one particular BH UCB is illustrated in Fig. 2.
In this example, the system enters the ultracompact phase via channel BH:02.
The system emerges from the second CE phase as a binary composed of a  
$1.2 \msun$ main sequence helium star and a heavy ONeMg WD companion of $\sim 1.4
\msun$ orbiting about their common center of mass with a period of about
1 hour. The helium star initially transfers material at a rate of $\sim 8 \times 10^{-8} 
\mpy$ and very quickly the accreting WD collapses to form a NS. 
In our present calculations we assume that
there is no natal kick associated with the AIC process. The MT rate initially exceeds
the Eddington limit for the NS accretor, but soon thereafter decreases and the NS 
accumulates the material efficiently. The evolution in this phase is similar to the evolution 
of a helium star - NS system as calculated by Savonije, de Kool, \& van den Heuvel 
(1986).  After about 30 Myrs into the MT phase, the NS
exceeds $2.0 \msun$ and a second AIC takes place, with the NS collapsing to form a BH.
As the MT continues, the He star loses most of its mass eventually  
becoming a degenerate low mass hybrid WD.  
The system becomes detached and the MT terminates.
Since the onset of ultracompact MT phase, the orbital period has decreased from 60
to about 20 minutes, as mass and angular momentum were lost from the system.
The calculations reveal that the MT phase was interrupted briefly as the donor became 
degenerate, with the subsequent evolution leading to the increase in the orbital 
separation as the donor expands. Although the MT rates are initially 
high, they very soon become sub-Eddington.  It should be pointed out that the detached phase is a 
result of our simplified treatment of the transition of the donor from a semi-degenerate
to a degenerate state.  However, the evolution of the system, especially at later 
times, is not significantly affected by this treatment. 
Eventually, the WD has been reduced to 0.09 $\msun$, marking the end
of the rapid evolution of the system. In an additional 1 Gyr the MT rate
slowly decreases to $\sim 10^{-12} \mpy$, and the period slowly
increases to $\sim 70$ minutes until the WD reaches the mass of 0.01
$M_\odot$ at which point our calculations were terminated. We note that although 
finite temperature effects of the WD on the MT have not been taken into account 
(cf., Deloye \& Bildsten 2003) this evolution should be indicative of NS and BH 
UCBs formed via this channel.  

In the above example the system becomes a transient X-ray source, when the MT 
rate falls below certain critical value (see Menou, Perna, \& Hernquist 1999; marked on Fig. 2). 
In our standard simulation we have assumed that all the material transferred  
during the transient stage is accreted by the compact star (NS or BH).  
However, it is possible that little material is accreted during the transient 
outburst stage. Had we assumed  that accretion was not effective 
during this stage in our example, 
the results would have been unchanged since the NS already had accreted sufficient 
matter to become a BH prior to the system entering the transient stage.

\subsection{Content of Current Population}

The population of NS/BH UCBs formed in the disk of our Galaxy 
at the current epoch (t=10 Gyrs) is listed in Table 2. 
Systems with NS accretors dominate the population (80\%), however
there is a non-negligible contribution of systems with BH accretors 
(20\%).  
The majority of the systems (both with NS and BH accretors) have  
WD companion donors.
As expected for systems with orbital periods less than 80 minutes the only
other donors found are low mass He-stars ($M \leq 2 \msun$). 
Many of the systems with WD donors have evolved through 
the phase when the donor was a He-star. 
However this phase is extremely short lived (by about 2 orders of magnitude
less than with WD donors, see Fig. 2), and therefore would be difficult to 
observe. BH UCBs may also evolve through a phase with a He-star donor, but for
these systems the lifetimes at this phase are even shorter (due to the lower mass of the He-star
donors) than in case of NS UCBs and are not detected in our 
simulated observed sample (see Table 2). 
For longer period binaries some BH UCBs (or heavy NS UCBs had we
raised the maximum NS mass to $2.5 \msun$ -- see below) may be fed by low 
mass MS donors.
Interestingly, we find that few NSs (below 10\%) in our population of 
UCBs are born in SN explosions, but are preferentially formed in 
AIC of heavy accreting ONeMg WD.  The case is more extreme for BH UCBs,
for which all the BHs are formed without a SN explosion through the AIC of a heavy 
accreting NS.

Systems with orbital periods less than 5 hours, but above 80 minutes are BHs
or heavy NSs with MS donors. These binaries evolve as proposed by Podsiadlowski, 
Rappaport, \& Pfahl (2002), starting as intermediate mass systems.
After a CE phase, the primary forms a NS in a SN
explosion.  A MS secondary of $2-4 \msun$ initiates MT at an orbital period of 
about 0.5 day. Prior to reaching a period of 5 hrs the NS mass exceeds $2 \msun$, and it
collapses to form a BH (this explains the absence of NS-MS systems with
periods shorter than 5 hrs in our sample of UCBs). 	
As the mass of the MS donor decreases, the orbital period decreases further  
near to the point at which the donor 
ceases nuclear burning in the core (corresponding to a mass lying 
between $0.08 \msun$ for a unevolved main sequence star to $0.35 \msun$ for 
an initially non degenerate helium core).   
Subsequently, the donor becomes a degenerate 
WD and the MT causes 
an increase of the orbital period (up to $\sim$ 60 min or longer) until the 
exhaustion of the donor (e.g., BH:06 formation channel).
Although these systems may constitute a tenth of BH UCBs, they give only a
small contribution to the entire NS/BH UCB population (less than a few
percent).

In order to assess the uncertainty of our predictions, we have calculated
several additional models with different assumptions on our input physics.
In one calculation we have relaxed the assumption of full accumulation 
during the transient stages, and have assumed that no material is accreted 
during these stages. 
The results of that calculation show only a slight change of the UCB
population, increasing the relative number of NS to 83.7\% and decreasing
the relative number of BH to 16.3\% of the total UCB population.

Yet another limiting factor for the formation of BH UCBs is the adopted
maximum NS mass.
The maximum mass has been estimated to be in the range of $1.8 - 2.3 \msun$
(see Akmal, Pandharipande, \& Ravenhall 1998) in comparison to an assumed maximum
of $2 \msun$ in our standard calculation.
However, most of the BHs in UCBs have rather low masses ($2-2.5 \msun$),
and an increase in the maximum mass has a significant effect on the BH population.
An increase of the maximum NS mass to $2.5 \msun$ enhances the population of
NS UCBs to 1.2, and reduces the expected number of BH UCBs to 0.1
($P_{orb} \leq 80$ min) or to 0.2 ($P_{orb} \leq  5$ hrs) of the standard
model number.

At least one more factor may affect the BH population. We have assumed that
NSs
may accrete all the transferred material up to the 
Eddington limit. This assumption led directly to the formation of BH
systems, 
since some of the accreting NSs were
able to accumulate  enough material to collapse to BHs
during the long lived MT episodes.
Had we relaxed 
that assumption, and allowed all the material to escape the systems, we would
end up with no BH UCBs in our population. However, neither the number of NS 
UCBs nor our conclusions about dominance of AIC systems in the current UCB 
population would be significantly altered.

Although there may be sufficient energy to eject the CE, the  
envelope in less evolved phases may not be sufficiently distended to significantly 
decelerate the spiral in process before the companion merges with the evolved core. 
Hence, those systems with less evolved donors which evolve into the CE phase 
may merge rather than survive. 
We estimate the reduction of systems entering into the UCB phase by assuming that 
donors which enter into the CE phase in the Hertzsprung gap do not 
survive. In this case, the number of NS UCBs is not significantly reduced, 
decreasing to 0.7 of the standard model number, however the BH UCBs would either 
vanish entirely ($P_{orb} 
\leq 80$ min) or are reduced to only 0.1 ($P_{orb} \leq 5$ hrs) of the 
standard model number. 

We have also calculated two models with different choices for the efficiency of
the common envelope ejection (in our standard model we assume $\alpha \times
\lambda = 1$; for details see Belczynski, Kalogera, \& Bulik 2002). 
An increase of the efficiency ($\alpha \times \lambda = 3$) barely changes the
results, however, a decrease of the efficiency ($\alpha \times \lambda = 0.1$) 
drastically reduces the number of the formed UCBs. In particular, the number of 
NS UCBs (71\%) 
is reduced by factor of $\sim 20$, while BH UCBs (29\%) 
is reduced 
by factor of $\sim
15$.  This is easily understood in the framework of the possible formation
scenarios of UCBs, many of which involve two episodes of CE. 
With a decreased efficiency the components of the binary progenitor systems
merge, as there is insufficient orbital energy to eject the CE, thus aborting
the formation of many UCBs.  We note that in this case, the numerical results 
reveal that all UCBs are formed via the AIC channel.

The formation
of ONeMg WDs plays an important role in our calculation since most 
of our UCBs are formed through the AIC of an ONeMg WD to a NS.
The lower mass threshold for ONeMg WD formation is found at an initial
stellar mass of $6.4 \msun$ (ONeMg WD mass of $1.2 \msun$) and the high 
end is at an initial mass of $8.0 \msun$ (ONeMg WD mass of $1.43  \msun$) 
for a solar metallicity 
within the framework of single stellar evolution models we use (Hurley et
al. 2000). Since the IMF in this mass range ($6-8 \msun$) is rather steep, 
most of ONeMg WD are formed with mass close to $1.2-1.3 \msun$.
Our models show that even if the ONeMg WD were formed at a reasonably lower
mass, some would be still pushed over the Chandrasekhar mass limit forming NSs and
UCBs, due to the sufficient mass reservoir in the progenitor systems. 

We have also chosen a different initial condition for our population of
primordial binaries. Instead of using correlated initial masses for two 
binary components (via a flat mass ratio distribution) we have calculated a 
model in which both masses are drawn independently.
Both components were taken within the same mass ranges as before (see \S\,2).
A broken power-law IMF for the stars in the Galactic disk which flattens out 
for low mass stars is used (Kroupa, Tout \& Gilmore 1993). 
For stars with initial masses $0.08<M<0.5 \msun$ the IMF 
slope is $\alpha_1=-1.3$, for $0.5<M<1 \msun$ the slope is $\alpha_2=-2.2$, 
while for $1<M<100\msun$ we use $\alpha_3=-2.7$. 
Such a prescription resulted in quite a drastic change of initial mass
ratios for our primordial binaries. After independently drawing two 
component masses for each binary, and recording the resulting mass ratios, 
we have obtained a distribution peaking below 0.1 and decreasing rapidly 
so there were almost no binaries with q $>$ 0.4. Such a distribution may be 
described by  $\propto q^{-2.7}$.
Due to the extreme mass ratio of most binaries, many of the previous progenitors
of UCBs 
do not survive the first episode of MT 
as a result of the 
occurrence of a dynamical instability, leading to merger in the ensuing CE.
We note an order of magnitude reduction of number of UCBs formed as compared 
to our standard model presented in Table 2.
This is very similar to findings of Han \& Podsiadlowski (2003) who noted
the order of magnitude decrease in the formation of Type Ia SNe progenitors 
with the independent choice of binary component masses.
There is also a significant reduction of relative number of systems with BH (6\%) 
as compared to the ones with NS (94\%) for $P_{orb} \leq 5$ hrs, with  
no BH UCBs formed for $P_{orb} \leq 80$ min.
The majority of NS systems are found with WD donors, while all BH systems are
formed 
with MS donors close to core hydrogen exhaustion.
However, even in this model our population is dominated by AIC systems
($\sim 90\%$) as compared to the UCBs with compact objects formed in SNe 
explosions ($\sim 10\%$).

\subsection{Predicted Number of Systems in the Galactic Field}

In this section we address the issue of the predicted absolute number of
UCBs in the Galactic field. 

Suggestions have been made (Clark 1975; Katz 1975; Verbunt \& Hut 1987) 
that dense stellar environments
and, in particular, globular clusters are very efficient in producing LMXBs. 
The hypothesis was put forward, that most if not all the systems were 
formed as the result of dynamical interactions in clusters, and then released 
from the clusters to populate the field (e.g., Grindlay 1984; White, Sarazin 
\& Kulkarni 2002).  
Different release mechanisms were proposed, including disruption of the clusters 
in the Galactic tidal field, ejection of systems due to 3- or 4-body 
interactions, and escape of the binaries with NS due to the gain of extra 
systemic velocities in SN explosions.

In the following we will show that the primordial field binaries are
sufficient to produce the  field NS UCBs (which are 
the subgroup of LMXB population).
Although the formation of such systems in globular clusters is possible,  
we point out that binary evolution with no dynamical
processes involved can account for the entire population of observed 
Galactic field UCBs. 

The calibration of the population synthesis results may be performed in several
ways. We choose to obtain absolute numbers using the observed star formation
rate in the disk of the Milky Way. However, for consistency calibration is also  
obtained  by comparison of our study to the measured rate of Type II 
and Ib/c SNe.
Galactic SFR have been estimated to lie in ranges of $1-3\ 
{\rm M}_\odot {\rm yr}^{-1}$ (Blitz 1997; Lacey \& Fall 1985) and $\sim
1-10\ {\rm M}_\odot {\rm yr}^{-1}$ (Gilmore 2001).
Cappellaro et al. (1999) estimated the rates of Type II SN and Type Ib/c
SN to be $1.86 \pm 0.35$ SNu and $0.14 \pm 0.07$ SNu for Sbc-d galaxies,
where 1 SNu corresponds to one SN per 100 yr and the estimates are
normalized to a blue luminosity of $10^{10}{L_\odot}^B$.  For an
estimated Galactic blue luminosity of about ${L_\odot}^B=2 \times
10^{10} L_\odot$ (van der Kruit 1987) we obtain 1.72 and 0.28 SNu for 
Type II and Ib/c SN respectively.  
 
Extending our study of stellar masses to the H burning limit ($0.08 M_\odot$)
with the use of Kroupa et al. (1993) broken power-law IMF  
the average (continuous) star formation rate which corresponds to the number 
of UCBs formed in our sample can be estimated. 
Comparison with the observed rate requires an upward revision for the
numbers presented in Table 2 by a factor of 100 
to account for the entire population within the Galaxy.
To be conservative, and not to overestimate the number of UCBs formed in
the field, we use the lower bound on the star formation in the disk of 
$1\ {\rm M}_\odot {\rm yr}^{-1}$. 
The X-ray duty cycle (DC) for transients sources is required, and we  
adopt  $DC \lesssim 1\%$ (Taam, King \& Ritter 2000). Only $\sim 1\%$ 
of our transient systems (listed in Table 2) have a chance to be observed 
in the outburst state, leading to a reduction by factor $\sim 33$.
We also note the possible further reductions of UCBs due to the largest uncertainties 
of stellar evolution, in particular CE evolution (factor of $\sim 20$) and the 
rather arbitrary choice of initial mass function (factor of $\sim 10$).
Combining all the above factors lead us to estimate the number of active 
UCBs in the field at the present time to be 7. 
However, we should understand 
this estimate as a lower limit, due to our conservative choice of models 
tending to reduce the predicted number of UCBs. 
Utilizing the combined observed rate of Type II and Ib/c SN for our 
calibration results in essentially the same number.

The number of observed confirmed field  UCBs hosting NS is 4, with 4 
strong candidates and a few additional systems potentially connected to that 
group. Moreover, 
there are observational uncertainties since many of the faint X-ray sources have no
orbital information, and some may yet contribute to the UCB population. Therefore, one may expect 
the number of the active field systems to be $\gtrsim 10$, which is 
consistent with our conservative estimate.

\section{Conclusions}

We have calculated the galactic population of short period ($P < 80$ min) 
ultracompact binary systems and shown that their formation follows from 
the evolutionary channels of progenitor binary systems characterized primarily
by stars of an intermediate mass range ($5 < M/\msun < 8$). Although UCBs exist in the 
dense cores of globular clusters, their population in the Galactic field 
does not necessarily require their production in these stellar systems. Our population
synthesis results reveal that UCB systems with BH accretors as well as NS 
accretors can be formed with the ratio of the former to the latter systems 
amounting to 1 - 20\% (accounting for a reasonable range of input parameters). 

An examination of the formation process of the compact objects in these systems 
reveals that, in contrast to previous investigations, the immediate progenitors of 
the majority of UCBs are ONeMg white dwarfs which undergo accretion induced collapse 
to a neutron star in response to the accretion of matter from hydrogen exhausted 
companions.  This new pathway results in 
the unexpected potential importance of the accretion induced collapse channel
for the majority (90\%) of the UCB systems.
Similarly,  
the BHs formed in these systems are not produced directly from the collapse of 
a massive star, but rather by the accretion induced collapse of a neutron star 
which has accreted sufficient mass from its companion.

With our adopted IMF slope (-2.7) ONeMg WDs (40\%) are formed almost as
frequently as NSs (60\%). 
Frequent MT episodes in close binaries 
further enhance the number of the WDs relative to NSs. 
However, the most important effect reducing the number of binaries 
hosting NSs, formed through direct core collapse, are SN explosions which tend to 
disrupt the binaries (due to the mass loss and natal kicks).
Considering the above arguments, and noting that ONeMg WDs are formed with a 
mass 
within several tenths of a $\msun$ 
to the Chandrasekhar limit (so that the mass accumulation during MT
required to convert them into NSs is not large) it is understandable why 
the NS UCB population is dominated by NSs formed through AIC. 
All BHs in our population of UCBs are formed through AIC.	
Since BHs formed through SN/core collapse originate from more massive
stars, there is only a very small probability that a progenitor system with a
low/intermediate mass companion survives a first MT episode, thus aborting 
this evolutionary channel for the formation of UCBs.

The donors in systems with NS accretors are restricted to 
helium WDs (60\%) and hybrid WDs (40\%) 
with a small contribution from low mass He
stars (1\%). In contrast, the donors 
of BH UCBs are hybrid WDs (95\%) and CO WDs (5\%). 
For longer period binaries ($P \leq 5$ hrs) there is a small but
significant (10\%) fraction of MS donors in BH UCBs (or NS UCBs had we
raised the maximum NS mass to $2.5 \msun$).

Those UCBs with NS accretors and He-rich donors are likely to give rise to 
Type I bursts as a result of a thermonuclear instability on the NS surface.  
Depending
upon the amount of residual hydrogen present on the white dwarf surface 
superbursts may be produced via thermonuclear burning of the underlying carbon 
layer (Strohmayer \& Brown 2002; see also Taam \& Picklum 1978) or due to the 
photodisintegration of heavy elements produced in the rapid proton capture
process (see Schatz, Bildsten, \& Cumming 2003).  Note, that only those 
WDs formed from the cores of hydrogen rich stars could be 
donors for the photodisintegration scenario, whereas the  He WDs 
formed from the evolution of He stars would lead to the carbon 
burning scenario for superbursts. For those rare UCBs with CO donors 
it is possible that they also ignite unstably and may produce a rare, but
energetic outburst on the NS. 

Many (40\%) of the NS-WD systems have evolved through the MT phase
with the He-star donor after the last CE phase. As a result these systems 
host hybrid low mass WDs, which may give a rise to a Ne-enriched
accretion flow later in the further evolution of these systems (Yungelson 
et al. 2002).

Only a small fraction (2.2\%) of UCBs are persistent X-ray sources, with 
X-ray luminosities of $L_x= 7 \times 10^{36} - 3 \times 10^{38}$ erg s$^{-1}$.
The majority of the UCBs are transient X-ray sources (97.8\%) with low MT
rates of $10^{-11} - 10^{-12} \mpy$. 
Without a reliable knowledge of the recurrence times we are unable to 
predict the precise number of active sources at the present time.
However, if only a small fraction of them are active (say 0.1) they still 
would dominate the current population of UCBs, both in number as well as 
in X-ray brightness with typical peak luminosities of 
$L_x= 4 - 8 \times 10^{38}$ erg s$^{-1}$ 
(assuming accretion at the Eddington rate during active states).

Out of 4 observed systems, 2 are transients, one is probably a
persistent source, and one is X-ray burster. Moreover, the majority host millisecond 
pulsars (3 out of 4) and their orbital periods span the range of 40 to 50 minutes.
Given the theoretical and observational uncertainties, these observations 
are not inconsistent with our results.
Specifically, more transient sources are predicted than 
persistent sources, however, about half of the systems (with a He rich donor) 
are expected to be X-ray bursters. Since most of
the NSs have accreted $\geq 0.1 \msun$, the NSs would likely have been spun up to millisecond 
spin periods.  Finally, most of these systems 
spend 90\% of their ultracompact lifetime after emerging
from a period minimum at orbital periods $\sim 40-60$ minutes, with 
very low mass ($0.01-0.1 \msun$) H-depleted degenerate donor stars.

\acknowledgements We would like to thank N. Ivanova, V. Kalogera, F. Rasio, 
A. Gurkan, J. Grindlay, A. King, and Gijs Nelemans for very useful 
discussions on this project.  
We also want to express our thanks to the anonymous referee who supplied us
with vital suggestions and comments.
This research was supported in part by the NSF under Grant No. AST-0200876 to 
RT and KBN grant 5PO3D01120 to KB.

\begin{deluxetable}{cccc}
\tablewidth{380pt}
\tablecaption{Ultracompact Binary Formation Channels}
\tablehead{ Formation & Efficiency\tablenotemark{a} & Efficiency & \\
            Channel & $P \leq 5\ hrs$ & 
            $P \leq 80\ min$ & Evolutionary History
\tablenotemark{b}  }
\startdata
NS:01 & 59.4\% & 60.9\% & CE1 CE2 MT2(NS-He/WD) \\
NS:02 & 10.4\% & 10.7\% & CE1 MT1 CE2 MT2(NS-He/WD) \\
NS:03 & 5.5\% & 5.6\% & CE1 SN1 CE2 MT2(NS-WD) \\  
NS:04 & 2.4\% & 2.5\% & CE1 MT1 SN1 CE2 MT2(NS-He/WD) \\
NS:05 & 0.8\% & 0.8\% & CE1 MT1 MT2 CE2 MT2(NS-WD) \\
NS:06 & 0.6\% & 0.6\% & CE1 MT2 CE2 MT2(NS-WD) \\ 
&&&\\
BH:01 & 7.3\% & 7.5\% & CE1 CE2 MT2(BH-WD) \\
BH:02 & 4.3\% & 4.4\% & CE1 MT1 CE2 MT2(BH-WD) \\
BH:03 & 3.7\% & 3.8\% & CE1 SN1 CE2 MT2(BH-He/WD) \\
BH:04 & 3.1\% & 3.1\% & CE1 MT1 SN1 CE2 MT2(BH-WD) \\ 
BH:05 & 1.8\% & 0.0\% & CE1 MT1 SN1 MT2(BH-MS/WD) \\
BH:06 & 0.6\% & 0.0\% & CE1 SN1 MT2(BH-MS/WD) \\ 
\enddata
\label{channels}
\tablenotetext{a}{Normalized to the NS/BH UCB population.}
\tablenotetext{b}{Sequences of different evolutionary phases:
common envelope CE, mass transfer MT, supernova explosion SN, followed by
the digit: 
1 stand for the primary (initially more massive component), and 2 for the 
secondary and these digits mark the donor in CE/MT events, and exploding
component in SN.
In parenthesis we list stellar types of components at current time during 
UCB phase: MS-- main sequence, He-- naked Helium star, WD-- white dwarf.
}
\end{deluxetable}

\begin{deluxetable}{crr}
\tablewidth{250pt}
\tablecaption{ Ultracompact Binaries in the Galactic Field}
\tablehead{ Type\tablenotemark{a}& $P \leq 5\ hrs$\tablenotemark{b}& $P \leq 80\ min$}
\startdata
NS accretor         &  79.1\% (388)                   &   81.2\% (388)\\
+ donor: WD/He/MS   &  78.6/0.6/0\%                   &  80.5/0.6/0\%\\
NS formed in SN:    &   8.0\%                         &   8.2\% \\
 &  &  \\
BH accretor         &   20.8\% (102)                   &   18.8\% (90)\\
+ donor: WD/He/MS   & 18.4/0.0/2.4\%                   &  18.8/0/0\%\\
NS formed in SN:    &   9.2\%                         &    6.9\%\\

\enddata
\label{numbers}
\tablenotetext{a}{
UCBs with NS and BH accretors are listed. For both groups the relative
occurrence frequency of given type of donor (He stand for naked helium star) is given.
The number of NSs formed in SN explosions (as opposed to formation through AIC) is
also given (all BHs are formed from NS through AIC, however some of these NS
are formed in SN, and their number is listed for the BH accretor UCB class).  
}
\tablenotetext{b}{Relative numbers of systems with orbital period less  
than $P$. In the parenthesis the actual number of systems formed in 
our simulation is given.}
\end{deluxetable}

\pagebreak

\psfig{file=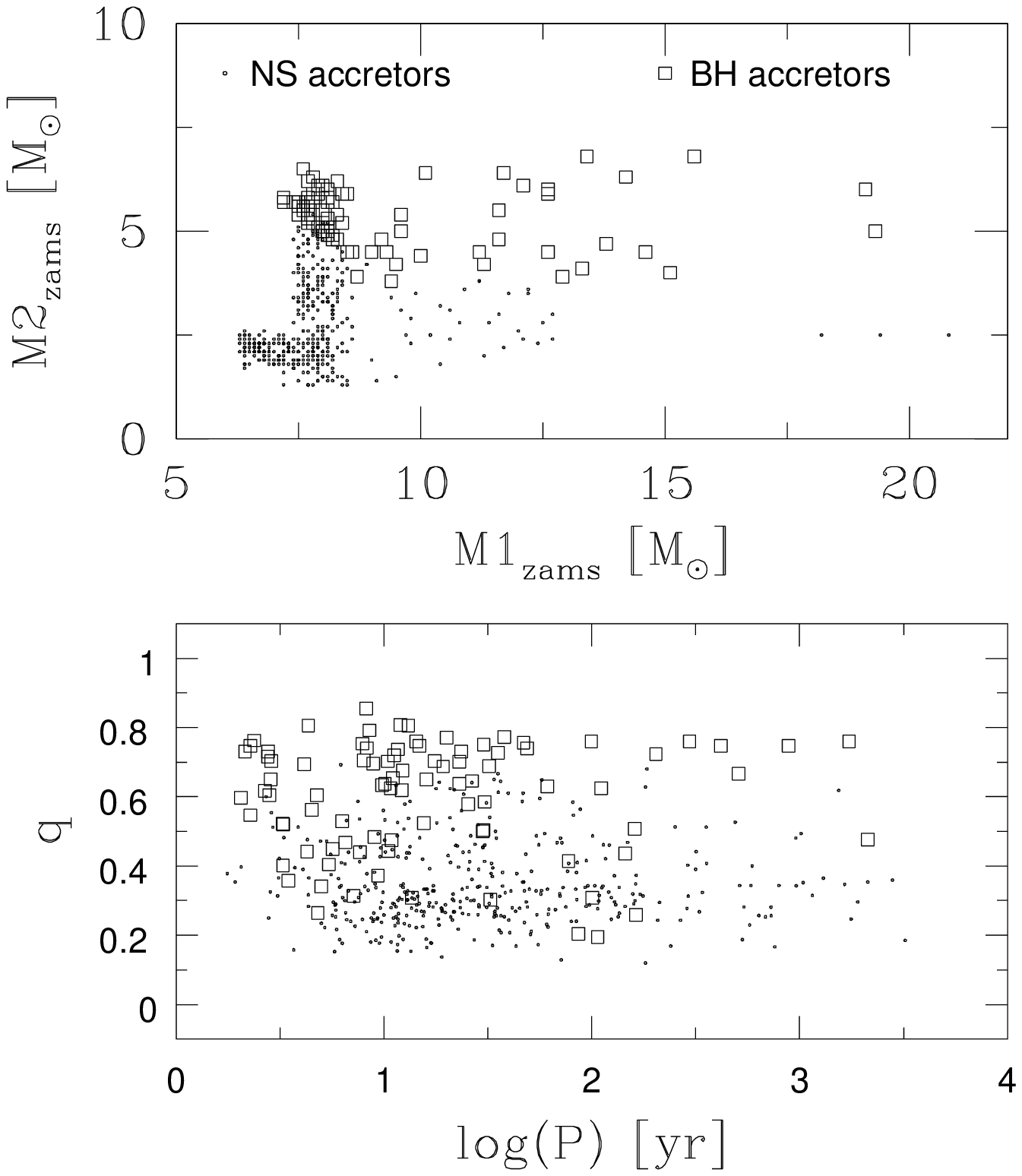,width=0.5\textwidth} 
\figcaption[]{
\footnotesize
Initial binary parameters of UCB progenitor systems.
Upper panel shows the correlation between the initial masses of the two 
binary components, where $M1_{\rm zams}$ and $M2_{\rm zams}$ denotes the  
initial mass of primary and secondary respectively.
The lower panel shows the mass ratios (secondary/primary) 
versus the initial orbital periods. 
}

\pagebreak

\psfig{file=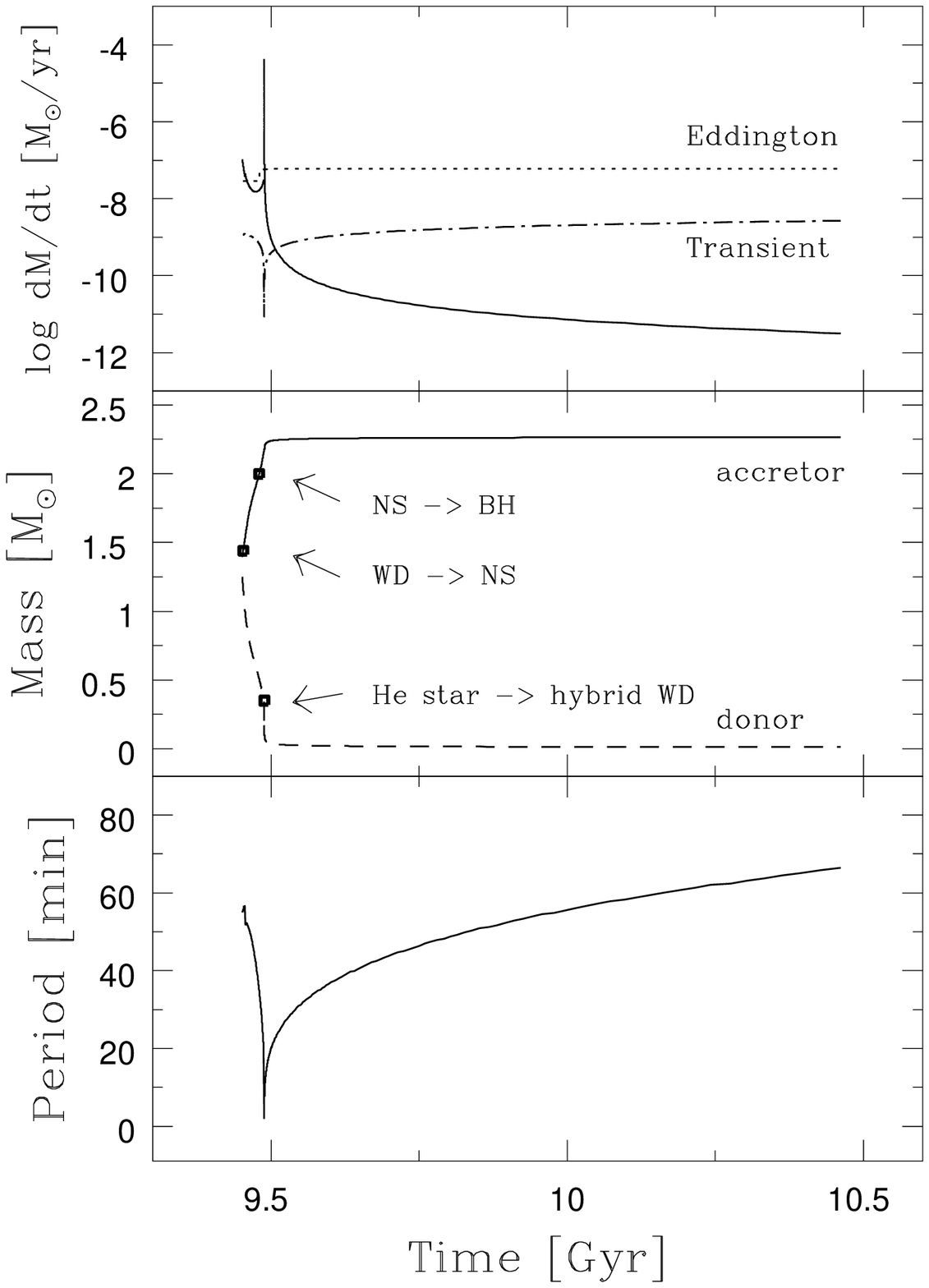,width=0.5\textwidth} 
\figcaption[]{
\footnotesize
An example evolution during the ultracompact phase.  The 
horizontal axis corresponds to the time after the formation of the Galactic disk.
The current time corresponds to 10 Gyrs at which
time the system is a WD-BH pair (with WD mass of 
$0.01 \msun$ and BH mass of $2.3 \msun$) at an orbital period of $\sim$ 60 min. 
The mass transfer rate is $7 \times 10^{-12} M_\odot yr^{-1}$ indicating
that the system is a transient X-ray source.
Upper panel: The temporal variation of the mass transfer rate is shown as a solid 
line.  The level of critical Eddington mass  accretion limit (dotted line) and 
the critical rate below which system exhibits transient behavior (dot-dashed line) 
are also shown. 
Middle panel: The variation of the binary component masses; for the accretor, the two
phases of AICs are marked; for the donor we mark the moment it became a WD. 
Lower panel: The orbital period evolution as a function of time.
}


\begin{references}

\reference{} Akmal, A., Pandharipande, V. R., \& Ravenhall, D. G. 1998, Phys. 
Rev. C, 58, 1804
\reference{} Arzoumanian, Z., Chernoff, D.\ F., \& Cordes, 
J.\ M.\ 2002, \apj, 568, 289
\reference{} Belczynski, K., Kalogera, V., \& Bulik, T.\ 2002,
\apj, 572, 407
\reference{}  Blitz, L.\ 1997, in `CO: Twenty-Five Years of  
Millimeter-Wave Spectroscopy'', eds.\ W.\ B.\ Latter, et al.\
(Kluwer Academic Publishers), p.\ 11
\reference{} Cappellaro, E., Evans, R., \& Turatto, M.\ 1999,
\aap, 351, 459
\reference{} Chakrabarty, D.\ 2003, KITP Workshop: The Physics of Ultracompact 
Stellar Binaries \\
(http://online.kitp.ucsb.edu/online/ultra\_c03/chakrabarty1/)
\reference{} Clark, G. W. 1975, \apj, 199, L143
\reference{} Deloye, C., \& Bildsten, L.\ 2003, \apj, submitted 
\reference{} Gilmore, G.\ 2001, Galaxy Disks and Disk
Galaxies, eds.\ J.G.\ Funes \& E.M.\ Corsini, San Francisco, ASP, p. 3
\reference{} Grindlay, J. E.\ 1984, Adv. Space. Res., 3, 19 
\reference{} Hachisu, I., Kato, M., \& Nomoto, K. 1996, \apj, 470, L97
\reference{} Han, Z., \& Podsiadlowski, Ph.\ 2003, astro-ph/0309618
\reference{} Hurley, J. R., Pols, O. R., \& Tout, C. A. 2000, \mnras, 315, 543
\reference{} Iben, I. Jr., \& Livio, M. 1993, \pasp, 105, 1373
\reference{} Ivanova, N., Belczynski, K., Kalogera, V., Rasio, F., \& 
Taam, R. E.\ 2003, \apj, 592, 475
\reference{} Ivanova, N., \& Taam, R. E. 2003, \apj, in press 
\reference{} Juett, A. M., Psaltis, D., \& Chakrabarty, D.\ 2001, \apj, 560,
L59
\reference{} Kato, M., \& Hachisu, I. 1999, \apj, 513, L41
\reference{} Katz, J. I. 1975, \nat, 253, 698
\reference{} Kawai, Y., Saio, H., \& Nomoto, K. 1987, \apj, 315, 229
\reference{} King, A. R., \& Begelman, M. C. 1999, \apj, 519, L169
\reference{} Kroupa, P., Tout, C.\ A., \& Gilmore, G.\ 1993,
\mnras, 262, 545 
\reference{} Lacey, C.\ G., \& Fall, S.\ M.\ 1985, \apj, 290, 154
\reference{} Li, X. D., \& van den Heuvel, E. P. J. 1997, \aap, 322, L9
\reference{} Menou, K., Perna, R., \& Hernquist, L. 2002, \apj, 564, L81
\reference{} Nelemans, G., Yungelson, L. R., \& Portegies Zwart, S.\ 2001,
\aap, 375, 890
\reference{} Nelson, L. A., \& Rappaport, S.\ 2003, astro-ph/0304374
\reference{} Podsiadlowski, Ph., Rappaport, S., \& Pfahl, E. D. 2002, \apj, 565, 1107
\reference{} Rasio, F., Pfahl, E., \& Rappaport, S.\ 2000, \apj, 532, L47
\reference{} Sandquist, E. L., Taam, R. E., \& Burkert, A. 2000, \apj, 533, 984
\reference{} Savonije, G. J., de Kool, M., \& van den Heuvel, E. P. J.\
1986, \aap, 155, 51 
\reference{} Schatz, H., Bildsten, L., \& Cumming, A. 2003, \apj, 583, L87
\reference{} Strohmayer, T. E., \& Brown, E. F. 2002, \apj, 566, 1045
\reference{} Taam, R. E., \& Picklum, R. E. 1978, \apj, 224, 210
\reference{} Taam, R. E., \& Sandquist, E. L. 2000, \araa, 38, 113
\reference{} Taam, R. E., \& van den Heuvel, E. P. J. 1986, \apj, 305, 235
\reference{} Tutukov, A.V., Fedorova, A.V., Ergma, E.V., \& Yungelson, 
L.R.\ 1987, Soviet Astronomy Letters, 13, 328
\reference{} van der Kruit, P.\ C.\ 1987, in ``The Galaxy'', eds.\ G.\   
Gilmore \& B.\ Carswell, Dordrecht: Reidel, p.\ 27
\reference{} Verbunt, F., \& Hut, P. 1987, The Origin and Evolution of Neutron Stars, 
eds. Helfand, D. J., \& Huang, J. H., (Dordrecht, Holland: Reidel), p. 187
\reference{} Warner, B. 1995, Cataclysmic Variable Stars, 
(Cambridge: Cambridge University Press)
\reference{} Webbink, R. F. 1992, X-Ray Binaries and Recycled Pulsars, ed., 
E. P. J. van den Heuvel \& S. A. Rappaport (Dordrecht: Kluwer), 269
\reference{} White, R. E. III, Sarazin, C. L., \& Kulkarni, S. R.\ 2002,
\apj, 571, L23
\reference{} Yungelson, L. R., Nelemans, G., \& van den Heuvel, E. P. J. 2002, 
\aap, 388, 546
\end{references}
\end{document}